\documentclass[a4paper,11pt,nohyper]{JHEP3}

\usepackage{graphicx}
\usepackage{multirow}

\def\a{\alpha}

\def\c{\gamma}
\def\d{\delta}
\def\e{\epsilon}           
\def\f{\phi}               
\def\g{\gamma}

\def\i{\iota}

\def\k{\kappa}                    

\def\n{\nu}
\def\o{\omega}  \def\w{\omega}
\def\t{\tau}
\def\u{\upsilon}
\def\x{\xi}

\def\D{\Delta}

\def\G{\Gamma}

\def\L{\Lambda}
\def\O{\Omega}  \def\W{\Omega}
\def\P{\Pi}
\def\Q{\Theta}
\def\S{\Sigma}
\def\U{\Upsilon}
\def\X{\Xi}
\def\del{\partial}              
\let\a=\alpha  \let\g=\gamma \let\d=\delta \let\e=\epsilon
   \let\i=\iota \let\k=\kappa
  \let\n=\nu \let\x=\xi 
 \let\t=\tau  \let\f=\phi \let\c=\chi \let\y=\psi
\let\w=\omega  \let\D=\Delta \let\Q=\Theta \let\L=\Lambda
\let\X=\Xi  \let\U=\Upsilon  \let\Y=\Psi
\let\C=\Chi \let\W=\Omega   \let\G=\Gamma

\def\nn{\nonumber} \def\bd{\begin{document}} \def\ed{\end{document}}
\def\ds{\documentstyle} \let\fr=\frac \let\bl=\bigl \let\br=\bigr
\let\Br=\Bigr \let\Bl=\Bigl
\let\bm=\bibitem
\let\na=\nabla
\let\pa=\partial \let\ov=\overline
\newcommand{\be}{\begin{equation}}
\newcommand{\ee}{\end{equation}}
\def\ba{\begin{array}}
\def\ea{\end{array}}
\def\ft#1#2{{\textstyle{{\scriptstyle #1}\over {\scriptstyle #2}}}}
\def\fft#1#2{{#1 \over #2}}
\def\del{\partial}
\def\sst#1{{\scriptscriptstyle #1}}
 \def\oneone{\rlap 1\mkern4mu{\rm l}}
\def\ie{{\it i.e.\ }}
\def\via{{\it via}}
\def\semi{{\ltimes}}
\def\str{{\rm str}}
\def\Dm{{{D_{\sst{max}}}}}
\def\vac{ \left | 0 \right \rangle }
\def\kvac{ \left | k \right \rangle }

\def\sp{\; \; \;}

\def\bol{ \left | B (p^+) \right \rangle}
\def\bo1{ \left | B^0 (p^+) \right \rangle}

\def\bolt{ \left | B (p^+) \right \rangle_{\t}}

\def\boxl{ \left | B (x^-) \right \rangle}
\newcommand{\bea}{\begin{eqnarray}}
\newcommand{\eea}{\end{eqnarray}}

\def\<{ \langle }
\def\>{ \rangle }

\def\S{\Sigma}

\usepackage{amsmath,latexsym,amssymb,slashed}
\usepackage{graphicx}
\usepackage{psfrag}
\usepackage{subfigure}

\renewcommand{\floatpagefraction}{0.6}
\renewcommand{\textfraction}{0.2}

\newcommand\ca{\mathcal{A}}
\newcommand\vp{\varphi}
\newcommand\beal{\begin{align}}
\newcommand\bbone{\ensuremath{\mathbbm{1}}}

\newcommand{\eq}[1]{\begin{equation}#1\end{equation}}
\newcommand{\spl}[1]{\begin{split}#1\end{split}}
\newcommand{\al}[1]{\begin{align}#1\end{align}}
\newcommand{\subeq}[1]{\begin{subequations}#1\end{subequations}}

\newcommand{\arXividhepth}[1]{\href{http://arxiv.org/abs/#1}arXiv:{\tt #1} [hep-th]}
\newcommand{\arXividother}[2]{\href{http://arxiv.org/abs/#1}arXiv:{\tt #1} [#2]}

\newcommand{\bg}[1]{\hat{#1}}
\newcommand{\wj}{\widetilde{J}}
\newcommand{\reo}{\mathrm{Re}~\!\omega}
\newcommand{\imo}{\mathrm{Im}~\!\omega}
\newcommand{\ads}{AdS_4}
\newcommand{\mcal}{\mathcal{M}}
\newcommand{\ccal}{\mathcal{C}}
\newcommand{\ncal}{\mathcal{N}}
\newcommand{\boxedeq}[1]{
\begin{equation}
\fbox{
\rule[0.7cm]{0pt}{0pt}
$#1$
\rule[-0.45cm]{0pt}{0pt}
}
\end{equation}
}

\def\d{\text{d}}

\def\slashchar#1{\setbox0=\hbox{$#1$}           
\dimen0=\wd0                                 
\setbox1=\hbox{/} \dimen1=\wd1               
\ifdim\dimen0>\dimen1                        
\rlap{\hbox to \dimen0{\hfil/\hfil}}      
#1                                        
\else                                        
\rlap{\hbox to \dimen1{\hfil$#1$\hfil}}   
/                                         
\fi}

\def\Re           {{\rm Re\hskip0.1em}}
\def\Im           {{\rm Im\hskip0.1em}}

\newcommand{\E}{\text{\tiny E}}

\newcommand{\tV}{{\widetilde{V}}}
\newcommand{\tH}{{\tilde{h}}}
\newcommand{\tm}{{{m}}}
\newcommand{\tmu}{{\tilde{\mu}}}
\newcommand{\trho}{{\tilde{\rho}}}
\newcommand{\tv}{{\tilde{v}}}
\newcommand{\calo}{\mbox{${\cal O}$}}
\newcommand{\cala}{\mbox{${\cal A}$}}
\newcommand{\dd}{\mathrm{d}}
\newcommand{\ra}{\rightarrow}
\newcommand{\calv}{\mbox{${\cal V}$}}
\newcommand{\calh}{\mbox{${\cal H}$}}
\newcommand{\calm}{\mbox{${\cal M}$}}
\newcommand{\abs}[1]{\left| #1 \right|}
\newcommand{\zetaa}{{\psi}}
\newcommand{\tr}{{\rm tr}\,}

\newcommand{\ky}[1]{{\color{blue}{#1}}}

\title{Non-conformal entanglement entropy}

\author{Marika Taylor and William Woodhead  \\

Mathematical Sciences and STAG Research Centre, University of Southampton, \\
Highfield, Southampton, SO17 1BJ, UK.

\bigskip
 E-mail:
 \email{m.m.taylor@soton.ac.uk; w.woodhead@soton.ac.uk}}

\abstract{We explore the behaviour of renormalized entanglement entropy in a variety of holographic models: non-conformal branes; the Witten model for QCD; UV conformal RG flows driven by explicit and spontaneous symmetry breaking and Schr\"{o}dinger geometries. Focussing on slab entangling regions, we find that the renormalized entanglement entropy captures features of the previously defined entropic c-function but also captures deep IR behaviour that is not seen by the c-function. In particular, in theories with symmetry breaking, the renormalized entanglement entropy saturates for large entangling regions to values that are controlled by the symmetry breaking parameters.

} 

\newcommand{\bx}{\ensuremath{{\vec{x}}}}
\newcommand{\bk}{\ensuremath{{\vec{k}}}}
\newcommand{\bM}{\mathbf{M}}
\newcommand{\tps}[2]{\texorpdfstring{#1}{#2}}
\newcommand{\fpq}[2]{\ensuremath{{}_{#1} F_{#2}}}
\newcommand{\tfpq}[2]{\ensuremath{{}_{#1} \tilde{F}_{#2}}}
\newcommand{\tG}{\ensuremath{{\tilde{G}}}}

\begin{document}

\newcommand{\td}{\tilde}
 \newcommand{\bc}{\begin{center}}
 \newcommand{\ec}{\end{center}}
 \newcommand{\bfr}{\begin{flushright}}
 \newcommand{\efr}{\end{flushright}}
 \newcommand{\bfl}{\begin{flushleft}}
 \newcommand{\efl}{\end{flushleft}}
 \newcommand{\bt}{\begin{tabular}}
 \newcommand{\et}{\end{tabular}}

\section{Introduction}

Entanglement entropy is widely used in condensed matter physics, quantum information theory and, more recently, in high energy physics and black holes. Consider a reduced density matrix $\rho_{A}$, obtained from tracing out certain degrees of freedom from a quantum system. The associated entanglement entropy is then the von Neumann entropy:
\be
S = - {\rm Tr} \left ( \rho_A \ln \rho_A  \right ).
\ee
Throughout this paper we will be interested in the case for which a quantum system is subdivided into two, via partitioning space. In such a case $A$ is a spatial region, with boundary $\partial A$. 

The entanglement entropy characterizes the nature of the quantum state of a system. For example, in the ground state of a quantum critical system in $D$ spatial dimensions:
\be
S = c_{1-D} \frac{{\rm Area} (\partial A)}{\epsilon^{D-1}} + \cdots +  c_{0} \ln (R/\epsilon) + \tilde{c}_0, \label{cft}
\ee
where $c_{1-D}$, $c_0$ and $\tilde{c}_0$ are dimensionless; $R$ is a characteristic scale of the region $A$ and $\epsilon$ is an UV cutoff. Logarithmic terms arise when $D$ is odd, and their coefficients are related to the $a$ anomalies of the stress energy tensor. More generally, the famous area law leading term characterizes the ground state of a system and can be used to test trial ground state wavefunctions.  
Entanglement entropy can also be used to distinguish between different phases of a system, such as the confining/deconfining phase transition \cite{Klebanov:2007ws}. 

Continuum quantum field theory (with a cutoff) is often used as a tool to describe discrete condensed matter systems. In this context, the cutoff appearing in \eqref{cft} is related to the underlying physical lattice scale in the discrete system and the coefficients of power law terms such as $c_{1-D}$ capture the leading physical contributions to the entanglement entropy.  From a quantum field theory perspective, the expansion in \eqref{cft} implicitly assumes the use of a direct energy cutoff as a regulator. Different methods of regularisation result in different regulated divergences and thus the power law divergences are often called non-universal. By contrast logarithmic divergences are often denoted as universal as their coefficients are related to the anomalies of the theory.

In even spatial dimensions, the logarithmic term in \eqref{cft} is absent but the constant term $\tilde{c}_0$ is believed to be related to the number of degrees of freedom of the system. However, $\tilde{c}_0$ is manifestly dependent on the choice of the cutoff. In two spatial dimensions, if
\be
S = c_{-1} \frac{R}{\epsilon} + \tilde{c}_0 \label{d2}
\ee
for a spatial region with boundary of length $R$, then changing the cutoff as
\be
\epsilon \rightarrow \epsilon' \left (1 + \alpha \frac{\epsilon'}{R} + \cdots  \right )
\ee
for any choice of the dimensionless constant $\alpha$ gives 
\be
S = c_{-1} \frac{R}{\epsilon'} +  \left (\tilde{c}_0 - \alpha c_{-1} \right ),
\ee
so the constant term in the entanglement entropy clearly depends on the choice of regulator. 

If one is interested in isolating finite contributions to the entanglement entropy, one can evade the issue of regulator dependence. For example, if the entangling region is a strip of width $L$ and regulated length $R \gg L$, then the divergent contributions in \eqref{d2} cannot, by locality of the quantum field theory, depend on the width of slab, so
\be
c(L) = \frac{\partial S}{\partial L} = \frac{\partial \tilde{c}_0}{\partial L}
\ee
is finite as $\epsilon \rightarrow 0$ \cite{Casini:2006hu,Nishioka:2006gr,Calabrese:2009qy}. However, such an approach has several drawbacks. The regularisation is specific to the shape of the geometry (a slab) and a modified prescription is needed for curved entangling region boundaries such as spheres, for which the scale of the entangling region is related to the local curvature of the entangling region boundary (see proposals in \cite{Liu:2012eea}). Any such prescription depends explicitly on the UV behaviour of the theory. More generally, extraction of finite terms by differentiation obscures scheme dependence:  there is no connection with the renormalization scheme used for other QFT quantities such as the partition function and correlation functions. 

From a quantum field theory perspective, as opposed to a condensed matter perspective, it is very unnatural to work with a regulated rather than a renormalized quantity. In previous papers \cite{Taylor:2016aoi,Taylor:2016kic}, we introduced a systematic renormalization procedure for entanglement entropy, in which the counterterms are inherited directly from the partition function counterterms.  As we review in section \ref{two}, such renormalization guarantees that the counterterms depend only on the quantum field theory sources (non-normalizable modes in holographic gravity realisations) and not on the state of the quantum field theory (normalizable modes in holographic gravity realisations).

The renormalized entanglement entropy $S_{\rm ren}$ expressed as a function of a characteristic scale of the entangling region implicitly captures the behaviour of the theory under an RG flow: small entangling regions probe the UV of the theory, while larger regions probe the IR. In this paper we will establish how these finite contributions to entanglement entropy behave in a variety of theories, using holographic models. 

\bigskip

The plan of this paper is as follows. In section \ref{two} we review the definition of renormalized entanglement entropy introduced in \cite{Taylor:2016aoi}. In section \ref{three} we calculate the renormalized entanglement entropy for a slab region in anti-de Sitter (in general dimensions). The latter is relevant for the non-conformal branes discussed in section \ref{four}, as the latter can be viewed as dimensional reductions of anti-de Sitter theories in general dimensions. In section \ref{four} we also compute the renormalized entanglement entropy for a slab region in the Witten holographic model for QCD. Section \ref{five} explores renormalized entanglement entropy for operator and driven holographic RG flows, which are UV conformal. In section \ref{six} we consider renormalized entanglement entropy in holographic Schr\"{o}dinger geometries. 
In section \ref{seven} we summarise the main features of the renormalized entanglement entropy, using both our holographic results and earlier perturbative/lattice calculations. We conclude in section \ref{eight}.

\section{Renormalized entanglement entropy} \label{two}

Entanglement entropy is usually calculated using the replica trick. The R\'{e}nyi entropies are defined as 
\begin{equation}
S_n = \frac{1}{(1-n)} \left ( \log Z(n) - n \log Z(1) \right ) \label{nc:renyi}
\end{equation}
where $Z(1)$ is the partition function and $Z(n)$ is the partition function on the replica space obtained by gluing $n$ copies of the geometry together along
the boundary of the entangling region. The entanglement entropy is obtained as the limit
\begin{equation}
S = \lim_{n \rightarrow 1} S_n. \label{nc:ee}
\end{equation}
Note that this limit implicitly assumes that the R\'{e}nyi entropies are analytic in $n$. 

Both sides of~\eqref{nc:renyi} are UV divergent. In a local quantum field theory, the UV divergences of $\log Z(n)$ cancel with those of $n \log Z(1)$ except at the boundary of the entangling region; therefore the $UV$ divergences of $S(n)$ scale with the area of this boundary. 

We can formally define the renormalized entanglement entropy as \cite{Taylor:2016aoi}
\begin{equation}
S^{\rm ren}_{n} = \frac{1}{(1-n)} \left ( \log Z^{\rm ren}(n) - n \log Z^{\rm ren}(1) \right ) \label{nc:renyi2}
\end{equation}
with $S^{\rm ren} = S^{\rm ren}_1$. Here the renormalized partition functions are defined with any suitable choice of renormalization scheme.  

The replica space matches the original space, except at the boundary of the entangling region where there is a conical singularity. To define the renormalization on the replica space it is therefore natural to work within a renormalization method that works for generic curvature backgrounds for the quantum field theory. 

\subsection{Direct cutoff: field theory}

Consider for example a Euclidean free massive scalar field theory on a background geometry ${\cal M}$ of dimension $d$ and let $Z(1)$ be the partition function in the ground state. Using locality of the quantum field theory and dimensional analysis, the UV divergences in the partition function behave as 
\begin{equation}
\log Z(1) =  a_d V_d \Lambda^d + a_{d-2} m^2 V_d \Lambda^{d-2} + b_{d-2} \Lambda^{d-2} \int_{\cal M} d^d x \sqrt{h} {\cal R} + \cdots
\end{equation}
where $\Lambda$ is the UV cutoff, $V_d$ is the volume of the background (Euclidean) geometry, $m^2$ is the mass and ${\cal R}  $ is the Ricci scalar. The coefficients $(a_d, a_{d-2},b_{d-2},\cdots)$ are dimensionless and in the above expressions we ignore boundaries of ${\cal M}$. 

The divergences of the partition function on the replica space $Z(n)$ have exactly the same structure and coefficients. However, the curvature of the replica space has an additional term from the conical singularity \cite{Solodukhin:2008dh}
\begin{equation}
{\cal R}_n = {\cal R} + 4 \pi (n-1) \delta (\partial \Sigma) + {\cal O}(n-1)^2, \label{replica}
\end{equation}
where $\delta(\partial \Sigma)$ is localised on a constant time hypersurface, on the boundary of the entangling region. (Here and in what follows we consider only static situations.) Therefore, when we use the replica formula~\eqref{nc:ee} the leading divergences of the partition functions cancel so that the leading divergence in the entanglement entropy behaves as
\begin{equation}
S_{\rm reg}  = 4 \pi b_{d-2} \Lambda^{d-2} \int _{\partial \Sigma} d^{d-2} x \sqrt{\gamma} + \cdots
\end{equation}
Such a divergence can clearly be cancelled by the counterterm
\begin{equation}
S_{\rm ct} = -  4 \pi b_{d-2} \Lambda^{d-2} \int _{\partial \Sigma} d^{d-2} x \sqrt{\gamma},
\end{equation}
which is covariantly expressed in terms of the geometry of the entangling region. 

\subsection{Holographic renormalization}

In gauge-gravity duality, the defining relation is \cite{Gubser1998,Witten:1998qj}
\be
I_{E} = - \log Z,
\ee
where $I_E$ is the onshell action for the bulk theory dual to the field theory. In the supergravity limit this is given by the onshell Euclidean 
Einstein-Hilbert action together with appropriate matter terms i.e. 
\be
I_{E} = - \frac{1}{16 \pi G_{d+1}} \int_{ {\cal Y}_n} d^{d+1} x \sqrt{g} \left ( R + \cdots \right ) - \frac{1}{8 \pi G_{d+1}} \int_{\partial {\cal Y}_n}  d^{d} x \sqrt{h} \left ( K + \cdots \right ),
\ee
where the latter is the usual Gibbons-Hawking-York boundary term. The volume divergences of the bulk gravity action correspond to UV divergences of the dual quantum field theory; these divergences can be removed by appropriately covariant counterterms at the conformal boundary. 

For example, in the case of asymptotically locally anti-de Sitter solutions of Einstein gravity the action counterterms are
\be
I_{\rm ct} =  \frac{1}{8 \pi G_{d+1}} \int_{\partial {\cal Y}_n} d^{d} x \sqrt{h} \left ( (d-1) +  \frac{\cal R}{2 (d-2)} + \cdots \right )
\ee
where the ellipses denote terms of higher order in the curvature and logarithmic counterterms arise for $d$ even. 

Applying the replica formula to the bulk terms in the action, as discussed in \cite{Lewkowycz:2013nqa}, and using the analogue of \eqref{replica} for the bulk curvature, namely, 
\be
R_n = R + 4 \pi (n-1) \delta (\Sigma) 
\ee
gives the Ryu-Takayanagi functional \cite{Ryu:2006bv} for the entanglement functional:
\be
S = \frac{1}{4 G_{d+1}} \int_{\Sigma} \sqrt{h}.
\ee
Applying the replica formula to the counterterms gives
\be
S_{\rm ct} = - \frac{1}{4 (d-2) G_{d+1}} \int_{\partial \Sigma} \sqrt{\gamma} \left ( 1 + \cdots \right ), 
\ee
with the leading counterterm being proportional to the regulated area of the entangling surface boundary. 
Analogous expressions for higher derivative gravity and gravity coupled to scalars can be found in \cite{Taylor:2016aoi}.

Using a radial cutoff to regulate is perhaps the most geometrically natural way to renormalize the area of the minimal surface but it is not the only holographic renormalization scheme. Dimensional renormalization for holography was developed in \cite{Bzowski:2016kni} and this method could also be used to renormalize the holographic entanglement entropy. 

\section{AdS entanglement entropy in general dimensions} \label{three}

In this section we derive the renormalized entanglement entropy for a slab domain in Anti-de Sitter in 
general dimensions. This quantity in relevant to the non-conformal brane backgrounds discussed in the next section, as the latter can be understood
in terms of parent Anti-de Sitter theories, and also relevant for the Schr\"{o}dinger backgrounds discussed in section \ref{six}.  

Let us parameterise $AdS_{D+2}$ as
\begin{equation}
ds^2 = \frac{1}{\rho^2} \left ( d \rho^2 - dt^2 + dx \cdot dx_D \right ).
\end{equation}
The entangling functional is 
\begin{equation}
S = \frac{1}{4 G_{D+2}} \int_{\Sigma} d^D x \sqrt{h}
\end{equation}
We now consider an entangling region in the boundary of width $L$ in the $x$ direction, on a constant time hypersurface, longitudinal to the other $(D-1)$ coordinates $y^{\alpha}$. The bulk entangling surface is then specified by the hypersurface
$x(\rho)$ minimising 
\begin{equation}
S = \frac{1}{4 G_{D+2}} \int d^{D-1} y^{\alpha} \int \frac{d \rho}{\rho^D} \sqrt{1 + {(x')}^2} \label{nc:Dfunc}
\end{equation}
where $x' = \partial_{\rho} x$. The equation of motion admits the first integral
\begin{equation}
{(x')}^2 = \frac{\rho^{2D}}{(\rho_{0}^{2D} - \rho^{2D} )},
\end{equation}
where $\rho_{0}$ is the turning point of the surface, related to $L$ via
\begin{equation}
L = 2 \int_{0}^{\rho_0} \frac{\rho^D d \rho}{\sqrt{\rho_0^{2D} - \rho^{2D}}},
\end{equation}
or equivalently
\begin{equation}
L = 2 \rho_0 \int_0^1 \frac{x^D dx}{\sqrt{1 - x^{2D}}} = \rho_0 \left ( \frac{2 \sqrt{\pi} \Gamma \left ( \frac{1 +D}{2D} \right )}{\Gamma \left ( \frac{1}{2 D} \right )} \right ).
\end{equation}
The regulated onshell value of the entangling functional is then
\begin{equation}
S_{\rm reg} = \frac{V_y}{2 G_{D+2}} \int_{\epsilon}^{\rho_0} \frac{d \rho}{\rho^D \sqrt{1 - \frac{\rho^{2D}}{\rho_0^{2D}}}}
\end{equation}
where $V_y$ is the regulated volume of the $y^{\alpha}$ directions. For $D > 0$ the only contributing counterterm is the regulated area of the boundary
i.e.
\begin{equation}
S_{\rm ct} = - \frac{1}{4 (D-1) G_{D+2}} \int_{\partial \Sigma} d^{D-1} \sqrt{\tilde{h}}
\end{equation}
(where we assume that $D \neq 1$) and therefore 
\begin{equation}
S_{\rm ren} = \frac{V_y}{2 G_{D+2}} \left [ \int_{\epsilon}^{\rho_0} \frac{d \rho}{\rho^D \sqrt{1 - \frac{\rho^{2D}}{\rho_0^{2D}}}} - \frac{1}{(D-1) \epsilon^{D-1}} \right ],
\end{equation}
which can be rewritten in terms of dimensionless quantities as 
\begin{equation}
S_{\rm ren} = \frac{V_y}{2 G_{D+2} \rho_0^{D-1}} \left [ \int_{\tilde{\epsilon}}^{1} \frac{d x}{x^D \sqrt{1 - x^{2D}}} - \frac{1}{(D-1) \tilde{\epsilon}^{D-1}} \right ].
\end{equation}
This can be evaluated to give
\begin{equation}
S_{\rm ren} =  \frac{\sqrt{\pi} V_y}{4 D G_{D+2} \rho_0^{D-1}} \frac{\Gamma  \left ( \frac{1}{2 D} - \frac{1}{2} \right )}{\Gamma \left ( \frac{1}{2D} \right )} 
\end{equation}
and hence
\begin{equation}
S_{\rm ren} = - \frac{V_y}{4 (D-1) G L^{D-1}} {\left( \frac{ 2 \sqrt{\pi} \Gamma  \left ( \frac{1}{2 D} + \frac{1}{2} \right )}{\Gamma  \left ( \frac{1}{2 D} \right )} \right )}^D \label{nc:hol-co}
\end{equation}
As we discuss later, this quantity is closely related to the entropic $c$ function for slabs in anti-de Sitter computed in \cite{Nishioka:2006gr}. 
In the case of $D=1$ ($AdS_3$) the entangling functional is logarithmically divergent, and the renormalized entanglement entropy depends explicitly on the 
renormalization scale: for a single interval 
\begin{equation}
S_{\rm ren} = \frac{1}{2 G_3} \log \left ( \frac{2}{\mu} \right ), 
\end{equation}
where $\mu$ is the (dimensionless) renormalization scale.

\section{Non-conformal branes} \label{four}

In this section we will consider entangling surfaces in Dp-brane and fundamental string backgrounds. It is convenient to express these backgrounds in the so-called dual frame in ten dimensions as \cite{Boonstra:1998mp}
\begin{equation}
I_{10} =  - \frac{N^2}{{(2 \pi)}^7 \alpha'^4} \int d^{10}x \sqrt{G} N^{\gamma} e^{\gamma \phi} \left ( R(G) + \beta {(\partial \phi)}^2 - \frac{1}{2 (8-p)! N^2} |F_{8-p}|^2 \right )
\end{equation}
where the constants $(\beta,\gamma)$ are given below for Dp-branes and fundamental strings respectively. (Note that it is convenient to
express the field strength magnetically, so for $p < 3$ we use $F_{p+2} = \ast F_{8-p}$.) 

The field equations admit $AdS_{p+2} \times S^{8-p}$ solutions with a linear dilaton. The field equations following from the action above can be reduced over a sphere, truncating to 
a $(p+2)$-dimensional metric and scalar. The resulting action is then
\begin{equation}
I_{d+1} = - {\cal N} \int d^{d+1}x \sqrt{g} e^{\gamma \phi} \left ( R + \beta {(\partial \phi)}^2 + C \right ) \label{nc:Id1}
\end{equation}
where $d= p +1$ and the constants $({\cal N}, \beta, \gamma,C)$ depend on the type of brane under consideration. 

For Dp-branes 
\begin{align}
\gamma &= \frac{2 (p-3)}{(7-p)} & \beta &= \frac{4 (p-1)(p-4)}{{(7-p)}^2} \\
C &= \frac{2 (9-p)(7-p)}{{(5-p)}^2} & {\cal N} &= \delta_p N^{\frac{7-p}{5-p}} g_d^{2(p-3)/(5-p)} \nonumber
\end{align}
where 
\begin{equation}
\delta_p = \frac{2^{\frac{2(p-4)}{p-5}} {(5-p)}^{\frac{9-p}{p-5}} \pi^{\frac{p+1}{p-5}} {\Gamma\left(\frac{7-p}{2}\right)}^{\frac{p-7}{p-5}}}{\Gamma\left(\frac{9-p}{2}\right)}
\end{equation}
and $g_d^2$ is the dimensionful coupling of the dual field theory, which is related to the string coupling as
\begin{equation}
g_d^2 = g_s {(2 \pi)}^{p-2} {(\alpha')}^{\frac{p-3}{2}}.
\end{equation}
At any length scale $l$ there is an effective dimensionless coupling constant
\begin{equation}
g_{\rm eff}^2(l) = g_d^2 N l^{3-p}
\end{equation}
For the fundamental string
\begin{align}
\gamma &= \frac{2}{3} & \beta &= C = 0 \\
{\cal N} &= \frac{g_s N^{\frac{3}{2}} {(\alpha')}^{1/2}}{6 \sqrt{2}} \nonumber
\end{align}
and the dimensionful coupling is
\begin{equation}
g_f^2 = \frac{1}{2 \pi g_s^2 \alpha'}
\end{equation}
so
\begin{equation}
{\cal N} = \frac{N^{\frac{3}{2}}}{12 \sqrt{\pi} g_f}.  
\end{equation}

In all cases, the dual frame is chosen such that the equations of motion admit an $AdS_{d+1}$ solution:
\begin{align}
ds^2 &= \frac{1}{\rho^2} \left ( d \rho^2 +  dx \cdot dx_d \right ) \label{nc:con1} \\
e^{\phi} &= \rho^{2 \alpha} \nonumber
\end{align}
where the constant $\alpha$ again depends on the case of interest: for Dp-branes
\begin{equation}
\alpha = - \frac{(p-7)(p-3)}{4 (p-5)}
\end{equation}
while $\alpha = -3/4$ for fundamental strings. In general the equations admit an AdS solution with linear dilaton provided that the parameters are related as 
\begin{align}
  \alpha &= - \frac{\gamma}{2 (\gamma^2 - \beta)} \\
  C &= \frac{ (d (\gamma^2 - \beta) + \gamma^2)(d (\gamma^2 - \beta) + \beta)}{{(\gamma^2 - \beta)}^2}. \nonumber
\end{align}
For further discussion of this point, see \cite{Kanitscheider:2008kd}. 

The non-conformal branes are formally related to AdS gravity in the following way \cite{Kanitscheider:2009as}. Let us define a parameter $\sigma$ as
\begin{equation}
\sigma = \frac{d}{2} - \alpha \gamma.
\end{equation}
Now we consider $(2 \sigma + 1)$-dimensional gravity with cosmological constant $\Lambda  = - \sigma (2 \sigma - 1)$, so that the action is
\begin{equation}
I_{(2 \sigma + 1)} = - {\cal N}_{\rm AdS} \int d^{2 \sigma +1} x \sqrt{g_{2 \sigma +1}} \left ( R_{2 \sigma + 1} + 2 \sigma (2 \sigma -1 ) \right ).
\end{equation}
Reducing on a $(2 \sigma -d)$-dimensional torus with coordinates $z^a$ via a diagonal reduction ansatz
\begin{equation}
ds^2 = ds_{d+1}^2 (x) + \exp \left ( \frac{2 \gamma \phi}{ (2 \sigma -d)} \right ) dz^a dz_a \label{nc:red1}
\end{equation}
results in the action~\eqref{nc:Id1} where
\begin{equation}
{\cal N} = {\cal N}_{\rm AdS} V_{(2\sigma - d)},
\end{equation}
with $V_{(2 \sigma -d)}$ the volume of the compactification torus. 

\subsection{Entanglement functional and surfaces}

The entanglement functional follows from the replica trick: in the dual frame
\begin{equation}
S =  4 \pi {\cal N} \int _{\Sigma} d^{d-1}x \sqrt{h} e^{\gamma \phi} 
\end{equation}
The equations for the entangling surface can be expressed geometrically as 
\begin{equation}
K_m = \gamma \left ( \partial_m \phi - h^{ij} \partial_i X^p \partial_j X^n g_{mn} \partial_p \phi  \right )
\end{equation}
where $g_{mn}$ is the background metric, $h_{ij}$ is the induced metric on the entangling surface, $X^{m}(x^i)$ specifies the embedding of the entangling surface
into the background and $K_m$ are the associated traces of the extrinsic curvatures. 

The dual frame entanglement functional follows directly from the reduction of the pure gravity entanglement functional
\begin{equation}
S = 4 \pi {\cal N}_{\rm AdS} \int_{\Sigma_{2 \sigma -1}} d^{2 \sigma -1} x \sqrt{H}, 
\end{equation}
when one again uses the diagonal reduction ansatz~\eqref{nc:red1}, and assumes that the entangling surface wraps the torus and that the shape of the surface does not vary
along the torus directions. In the upstairs picture the entangling surface satisfies 
\begin{equation}
{\cal K}_{M} = 0,
\end{equation}
where the background metric is now denoted $g_{(2 \sigma + 1) MN}$ and ${\cal K}_{M}$ denotes the traces of the extrinsic curvatures. Thus, any AdS entangling surface
which factorises as $\Sigma_{2 \sigma -1} = T^{(2 \sigma -d)} \times \Sigma$ will give an entangling surface for non-conformal branes; moreover, the non-conformal brane surface will inherit its renormalized entanglement entropy from the upstairs entangling surface. 

As an example, let us consider slab entangling regions, characterised by a width $\Delta x = L$. The bulk entangling surface is specified as $x (\rho)$ and in the background~\eqref{nc:con1} the entangling functional is
\begin{equation}
S = 4 \pi {\cal N} V_y \int \frac{d \rho}{\rho^{2 \sigma -1}} \sqrt{1 + {(x')}^2},
\end{equation}
which is indeed precisely the functional obtained in~\eqref{nc:Dfunc}, identifying $D = (2 \sigma - 1)$.  The renormalized entanglement entropy can then be expressed as
\begin{equation}
S_{\rm ren} = - \frac{2 \pi {\cal N} V_y}{(\sigma -1) L^{2 (\sigma -1)}} {\left( \frac{2 \sqrt{\pi} \Gamma ( \frac{1}{2 (2 \sigma -1)} + \frac{1}{2} )}{\Gamma (\frac{1}{2(2 \sigma -1)})}\right)}^{2 \sigma -1}
\end{equation}

\bigskip

The renormalized entanglement entropy for a strip in the F1 background can be expressed as
\begin{equation}
S_{\rm ren} = - \frac {4 \pi^{\frac{3}{2}} {( \Gamma( \frac{3}{4} ) )}^2} { 3 {( \Gamma ( \frac{1}{4} ))}^2} \frac{N^2}{g_{\rm eff} (L)}
\end{equation}
where the effective coupling is expressed as $g^2_{\rm eff} (L) = g_f^2 N L^2$. The expression for the renormalized entanglement entropy of a strip in the D1 background is
analogous:
\begin{equation}
S_{\rm ren} = - \frac {4 \pi^{\frac{3}{2}} {( \Gamma( \frac{3}{4} ) )}^2} { 3 \sqrt{2} {( \Gamma ( \frac{1}{4} ))}^2} \frac{N^2}{g_{\rm eff} (L)}
\end{equation}

\subsection{Witten model}

The Witten \cite{Witten:1998zw} holographic model for YM$_4$ can be expressed in terms of the following six-dimensional background:
\bea
ds^2 &=& \frac{d \rho^2}{\rho^2 f(\rho)} + \frac{1}{\rho^2} \left ( - dt^2 + dx \cdot dx_3 + f(\rho) d \tau^2 \right ) \\
e^{\phi} &=& \frac{1}{\rho^{\frac{3}{2}}}, \nn
\eea
where
\be
f(\rho) = \left ( 1 - \frac{\rho^6}{\rho_{KK}^6} \right ).
\ee
Regularity of the geometry requires that the circle direction $\tau$ must have periodicity
\be
L_{\tau} = \frac{2 \pi}{3} \rho_{KK}.
\ee
This model originates from D4-branes wrapping the circle $\tau$ with anti-periodic boundary conditions for the fermions. which breaks the supersymmetry. At low energies
the model resembles a four-dimensional gauge theory, with the gauge coupling being $g^2 = g_5^2/L_{\tau}$. The gravity solution captures the behaviour of this theory in the limit
of large 't Hooft coupling $\lambda^2 = g^2 N \gg1$.

One of the main applications of this model is in the context of flavour physics: Sakai and Sugimoto \cite{Sakai:2004cn,Sakai:2005yt} introduced D8-branes wrapped around the $S^4$ on which the theory is reduced from ten to six dimensions. These D8-branes model chiral flavours in the dual gauge theory and the resulting Witten-Sakai-Sugimoto model has been used extensively as a simple holographic model of a non-supersymmetric gauge theory with flavours. 

The operator content of the dual theory captured by the metric and scalar field is the four-dimensional stress energy tensor $T_{ab}$, a scalar operator ${\cal O}_{\tau}$ corresponding to the component of the five-dimensional stress energy tensor $T_{\t\t}$ and the gluon operator ${\cal O}$ corresponding to the bulk scalar field. These operators satisfy a Ward identity \cite{Kanitscheider:2008kd}
\be
\langle T_{a}^a \rangle + \langle {\cal O}_{\tau} \rangle + \frac{1}{g^2} \langle {\cal O} \rangle = 0 
\ee
and their expectation values can be extracted from the above geometry. For example, the condensate of the gluon operator 
\be
\langle {\cal O} \rangle =  \frac{2^5 \pi^2}{3^7} \frac{\lambda^2 N}{L_{\tau}^4} 
\ee
and therefore $L_{\tau}$ controls the QCD scale of the theory. 

\bigskip

Next we can consider a slab entangling region, wrapping the circle direction $\tau$, characterised by a width $\Delta x = L$. Entanglement entropy in this theory 
was previously discussed in \cite{Klebanov:2007ws}, with the confinement transition being associated with a discontinuity in the derivative of the entanglement entropy with respect to $L$. 
The bare entanglement functional is
\be
S = 4 \pi {\cal N} V_2 L_{\tau} \int \frac{d \rho}{\rho^5} \sqrt{1 + f(\rho) (x')^2}
\ee
where $V_2$ is the volume of the two-dimensional cross-section of the slab. The entanglement can then be written as
\be
S_{\rm reg} =  8 \pi {\cal N} V_2 L_{\tau} \int_{\epsilon}^{\rho_{\ast}} \frac{d \rho}{\rho^5} \frac{\sqrt{f (\rho) \rho_\ast^{10}}}{\sqrt{f (\rho) \rho_{\ast}^{10} - f(\rho_{\ast}) \rho^{10}}}
\ee
where $\rho_{\ast}$ is the turning point of the surface, related to the width of the entangling region as
\be
L = 2 \int^{\epsilon}_{\rho_{\ast}} \frac{ d \rho}{\sqrt{f(\rho) \left ( \frac{\rho_{\ast}^{10}  f(\rho)}{\rho^{10} f(\rho_{\ast})} - 1 \right )}}.  
\ee
The entanglement entropy can be renormalized as before, with the counterterm contributions being
\be
S_{\rm ct} = - 2 \pi {\cal N} V_2 \frac{L_{\tau}}{\epsilon^4}. 
\ee 
For large entangling regions, the only possible entangling surface is the disconnected configuration, for which the renormalized entanglement entropy is 
\bea
S_{\rm ren} &=& 
8 \pi {\cal N} V_2 L_{\tau} \left ( \int^{\rho_{KK}}_{\epsilon} \frac{d\rho}{\rho^5} - \frac{1}{4 \epsilon^4} \right ) \\
&=& - \frac{4 \pi^2 {\cal N}}{3} \frac{V_2}{\rho_{KK}^3} \nn
\eea
For small entangling regions the condensate is negligible and the renormalized entanglement entropy is controlled by the conformal structure
\be
S_{\rm ren} \approx - \pi {\cal N} \left ( \frac{2 \sqrt{\pi} \Gamma(3/5)}{\Gamma(1/10)} \right )^5 \frac{V_2 L_{\tau}}{L^4}
\ee
The renormalized entanglement entropy is plotted in Figure~\ref{nc:fig-witten-plot}. As discussed in \cite{Klebanov:2007ws} there is a discontinuity in the derivative of the entanglement entropy for slab widths around $L \sim 0.4 \rho_{KK}$. For larger values of $L$ the entanglement entropy saturates at a constant value. 
\FIGURE[t]{\includegraphics*[width=\linewidth]{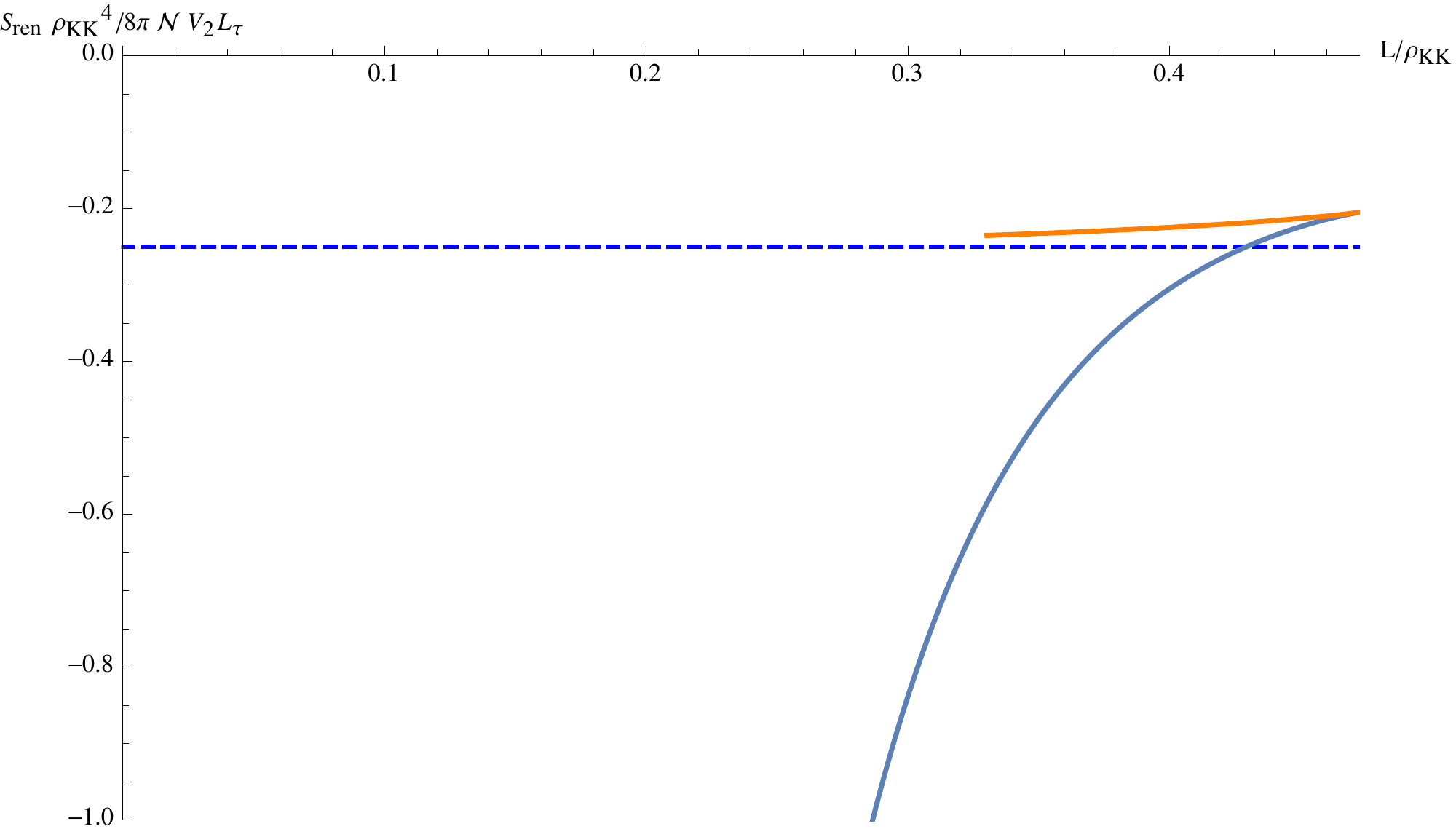}
  \caption{The renormalized entanglement entropy for the Witten model. The solid blue line and solid orange lines indicate the renormalized entropy for the two possible connected minimal surfaces.}\label{nc:fig-witten-plot}}

\section{Renormalized entanglement entropy for RG flows} \label{five} 

In this section we will consider holographic entanglement entropy in geometries dual to RG flows. We work in Euclidean signature with a bulk action
\begin{equation}
I = - \frac{1}{16 \pi G} \int d^{d+1} x \sqrt{g} \left ( R - \frac{1}{2} {(\partial \phi)}^2 + V(\phi) \right ). \label{nc:bact1}
\end{equation}
Holographic RG flows with flat radial slices can be expressed as 
\begin{equation}
ds^2 = dr^2 + \exp (2 A(r)) dx^i dx_i,
\end{equation}
where the warp factor $A(r)$ is related to a radial scalar field profile $\phi(r)$ via the equations of motion 
\begin{equation}
\frac{d^2 \phi}{d r^2 } + d \left ( \frac{ d A}{dr} \right ) \left ( \frac{d \phi}{dr} \right ) = - \frac{d V}{d \phi}  \qquad \qquad 
\frac{d^2 A}{d r^2} = - \frac{1}{2 (d-1) } {\left ( \frac{d \phi}{dr} \right )}^2. 
 \end{equation}
These equations can always be expressed as first order equations \cite{Freedman:2003ax}
\begin{equation}
\frac{dA}{d r} = W \qquad
\frac{d \phi}{d r} = - 2 (d-1) \frac{d W}{d \phi} \label{nc:1pde}
\end{equation}
where the (fake) superpotential $W(\phi)$ is related to the potential as
\begin{equation}
V = - {(d-1)}^2 \left ( 2 {\left ( \frac{d W}{d \phi} \right )}^2 - \frac{d}{d-1} W^2 \right ). 
\end{equation}
Near the conformal boundary the potential can be expanded in powers of the scalar field as 
\begin{equation}
V = d (d-1) - \frac{1}{2} m^2 \phi^2 + \cdots
\end{equation}
and hence the superpotential can be written as
\begin{equation}
W = 1 + \frac{(d- \Delta)}{4 (d-1)} \phi^2 + \cdots \label{nc:wexp}
\end{equation}
where $\Delta = d/2 + \sqrt{d^2 + 4m^2}/2$.  The higher order terms in the superpotential are not unique, as different choices are associated with different RG flows. 

Note that for flat domain walls, a single counterterm (in addition to the usual Gibbons-Hawking term) is sufficient
\begin{equation}
I_{\rm ct} = - \frac{(d-1)}{8 \pi G_4} \int d^d x \sqrt{\gamma}  W,
\end{equation}
although the derivation of the entanglement entropy counterterms requires knowledge of the counterterms for a curved background (since the replica
space is curved). 

\bigskip

The entanglement entropy for a slab region $\Delta x = L$ in the RG flow geometry is
\begin{equation}
S = \frac{V_y}{4 G} \int d r e^{(D-1) A(r)} \sqrt{1 + e^{2 A(r)} {(x')}^2},
\end{equation}
where $D$ is the number of spatial directions in the dual theory, 
$V_y$ is the regulated volume of the longitudinal directions and $x(r)$ defines the entangling surface. Then
\begin{equation}
L = 2 \int^{\infty}_{r_0} \frac{ e^{D A_0} dr}{e^{A(r)} \sqrt{ e^{2 D A(r)} - e^{2 D A_0}}},
\end{equation}
where at the turning point $r_0$ of the surface $A(r) = A_0$. The regulated onshell action is
\begin{equation}
S_{\rm reg} = \frac{V_y}{2 G}  \int_{r_0}^{\Lambda}  dr \frac{e^{(2 D-1) A(r)}} {\sqrt{e^{2 D A(r)} - e^{2 D A_0}}},
\end{equation}
with the cutoff being $r = \Lambda$. 

The entanglement entropy counterterms for RG flows driven by relevant deformations were discussed in~\cite{Taylor:2016aoi}, working perturbatively
in the deformation. Here we will analyse both spontaneous and explicit symmetry breaking, using exact supergravity solutions. 

\subsection{Spontaneous symmetry breaking: Coulomb branch of ${\cal N} = 4$ SYM}

In this section we consider
the case of VEV driven flow, i.e.\ spontaneous symmetry breaking. In such a situation, the scalar field has only normalizable modes and thus asymptotically
the scalar field behaves as
\begin{equation}
\phi \rightarrow \phi_{(0)} e^{- \Delta r} + \cdots
\end{equation}
where $\phi_{(0)}$ is related to the operator expectation value as 
\begin{equation}
\langle {\cal O} \rangle = - (2 \Delta - d) \phi_{(0)}. 
\end{equation}
From~\eqref{nc:1pde} and~\eqref{nc:wexp}, one can immediately read off the asymptotic form
of the warp factor:
\begin{equation}
A(r) = r - \frac{(d- \Delta)}{4 \Delta (d-1)} \phi_{(0)}^2  e^{-2 \Delta r} + \cdots 
\end{equation}
Substituting into the regulated action, we then obtain
\begin{equation}
S_{\rm reg} = \frac{V_y}{2 G}  \left ( \frac{ e^{(d-2) \Lambda}}{(d-2)} -  \frac{(d- \Delta)}{4 \Delta (d-1)(d -2 - 2 \Delta)} e^{(d -2  -2 \Delta) \Lambda} + \cdots \right ) 
\end{equation}
The second term vanishes as $\Lambda \rightarrow \infty$ for $\Delta > (d-2)/2$, and is logarithmically divergent for $\Delta = (d-2)/2$. (The latter case does not however arise holographically, as when the lower bound on the conformal dimension is saturated the operator automatically obeys free field equations.) Therefore, for VEV driven flows the only counterterm required is the regulated area of the boundary of the entangling surface:
\begin{equation}
S_{\rm ct} = - \frac{1}{4 (d-2) G} \int_{\partial \Sigma} d^{d-2} x \sqrt{h}.
\end{equation}
Note that one can derive the same result from the bulk action counterterms, using the replica trick; see below for the case of the Coulomb branch of ${\cal N} = 4$ SYM. Thus the renormalized entanglement entropy for slabs in VEV driven flows is 
\begin{equation}
S_{\rm ren} = \frac{V_y}{2 G} \left ( \int_{r_0}^{\Lambda}  dr \frac{e^{(2 D-1) A(r)}} {\sqrt{e^{2 D A(r)} - e^{2 D A_0}}} -  \frac{ e^{(d-2) \Lambda}}{(d-2)} \right ).
\end{equation}
Now let us consider the general structure of the renormalized entropy. In the vacuum of the conformal field theory, the renormalized entropy must behave as 
\begin{equation}
S_{\rm ren} = c_0 \frac{V_y}{G L^{D-1}}
\end{equation}
with $c_0$ a dimensionless constant on dimensional grounds: the entropy scales with the longitudinal volume $V_y$ and the width of the entangling region $L$ is the only 
other dimensionful scale in the problem. The value of $c_0$ in holographic theories is given in~\eqref{nc:hol-co}.

Now working perturbatively in the operator expectation value $\langle {\cal O} \rangle$ the renormalized entropy must behave as
\begin{equation}
S_{\rm ren} =  \frac{V_y}{G L^{D-1}} \left ( c_0  + c_1 \langle {\cal O} \rangle^2 L^{2 \Delta} + \cdots \right ) \label{nc:analytic}
\end{equation}
where $c_1$ is dimensionless and we work in a limit in which
\begin{equation}
\langle {\cal O} \rangle \ll \frac{1}{L^{\Delta}}
\end{equation}
i.e.\ the width of the entangling region is much smaller than the length scale set by the condensate. 

\subsubsection{Coulomb branch disk distribution}

We now analyse a specific example: the renormalized entanglement entropy of slab domains on the Coulomb branch of ${\cal N} = 4$ SYM. 

We consider the case of a disk distribution of branes preserving $SO(4) \times SO(2)$ symmetry, for which the equations of motion follow from~\eqref{nc:bact1}, with the superpotential being \cite{Freedman:1999gk}
\begin{equation}
W(\phi) = \frac{2}{3} \exp \left ( \frac{2 \phi}{\sqrt{6}} \right ) + \frac{1}{3} \exp \left ( - \frac{4 \phi}{\sqrt{6}} \right )
\end{equation}
The metric in five-dimensional gauged supergravity is then 
\begin{equation}
ds^2 = \lambda^2 w^2 \left ( \frac{d w^2}{\lambda^6 w^4} + dx \cdot dx \right )
\end{equation}
with 
\begin{equation}
\lambda^6 = \left ( 1 + \frac{\sigma^2}{w^2} \right ).
\end{equation}
Here the coordinate $w \rightarrow \infty$ at the conformal boundary and $\sigma$ characterises the expectation value of the dual scalar operator. The scalar field can be expressed by the relation
\begin{equation}
\sigma^2 \frac{ e^{ \frac{2}{\sqrt{6}} \phi}}{1 - e^{\sqrt{6} \phi}} =  \lambda^2 w^2. 
\end{equation}
Using the standard Fefferman-Graham coordinates near the conformal boundary:
\begin{align}
ds^2 &=  \frac{1}{\rho^2} d \rho^2 + \frac{1}{\rho^2} \left ( 1 - \frac{1}{18} \sigma^4 \rho^4 + \cdots   \right ) dx \cdot dx \\
\phi &= \frac{1}{\sqrt{6}} \sigma^2 \rho^2 + \cdots \nonumber
\end{align}
We can then read off the expectation values of the dual stress energy tensor and scalar operator, following~\cite{Bianchi:2001de,Bianchi:2001kw}:
\begin{equation}
\langle T_{ij} \rangle = 0 
\qquad
\langle {\cal O} \rangle = \frac{N^2}{\sqrt{6} \pi^2} \sigma^2
\end{equation}
where we use the standard relation between the Newton constant and the rank of the dual gauge theory:
\begin{equation} 
\frac{1}{16 \pi G_5} = \frac{N^2}{2\pi^2}.
\end{equation}
The vanishing of the dual stress energy tensor is required given the supersymmetry but careful holographic renormalization is required to derive this answer. 

\bigskip

The regulated entanglement entropy of a slab domain in this geometry can be written as 
\begin{equation}
S_{\rm reg} = \frac{V_2}{2 G_5} \int^{\Lambda}_{w_0} dw \lambda^3 w^3 \sqrt{ {(x')}^2 + \frac{1}{w^4 \lambda^6}}.
\end{equation}
Using the first integral of the equations of motion the width of the entangling region can be expressed in terms of the turning point of the surface $w_0$ as
\begin{equation}
L = 2 \int^{\infty}_{w_0} \frac{ c dw}{w^2 \lambda^3 \sqrt{\lambda^6 w^6 - c^2}}
\end{equation}
where $c$ is an integration constant and $w_0$ satisfies
\begin{equation}
w_0^4 ( \sigma^2 + w_0^2) = c^2. 
\end{equation}
The regulated entanglement entropy is then 
\begin{equation}
S_{\rm reg} = \frac{V_2}{2 G_5} \int_{w_0}^{\Lambda} dw \frac{ w^3 \sqrt{w^2 + \sigma^2}}{\sqrt{w^4 ( w^2 + \sigma^2 ) - c^2}}
\end{equation}
and the required counterterm is expressed in terms of the regulated area of the boundary of the entangling surface i.e.\ there are counterterm contributions
\begin{equation}
S_{\rm ct} = - \frac{V_2}{8 G_5} \Lambda^2 {\left ( 1 + \frac{\sigma^2}{\Lambda^2} \right )}^{\frac{1}{3}}
\end{equation}
at each side of the slab. (The total contribution is therefore twice this value.) Note that the counterterms in this case clearly contribute both divergent and finite parts: expanding in powers of the cutoff $\Lambda$
\begin{equation}
S_{\rm ct} = - \frac{V_2}{8 G_5} \Lambda^2 - \frac{V_2}{24 G_5} \sigma^2 + \cdots 
\end{equation}
It is then convenient to write the entanglement entropy  in terms of dimensionless quantities as 
\begin{equation}
S_{\rm ren} = \frac{V_2 \sigma^2}{4 G_5}  \lim_{\tilde{\Lambda} \rightarrow \infty} \left ( \int_{y_0}^{\tilde{\Lambda} } dy \frac{ y \sqrt{y+1}}{\sqrt{y^2 ( y + 1 ) - y_0^2 (y_0+1)}} - \tilde{\Lambda} - \frac{1}{3} \right ),
\end{equation}
where $\tilde{\Lambda}$ is a rescaled dimensionless cutoff. Implicitly this expression assumes that $\sigma^2\neq\nobreak 0$ and $y_0$ is the turning point of the surface. Then
\begin{equation}
\sigma L = y_0 \sqrt{y_0 + 1}  \int_{y_0}^{\infty} \frac{dy}{y \sqrt{(y+1)} \sqrt{y^2 (y+1) - y_0^2 (y_0 + 1)}}
\end{equation}
These integrals can be computed numerically. There is a maximal value of $L$ (for fixed $\sigma$) for which a connected entangling surface exists: the critical value of $L$ is such that 
\begin{equation}
\sigma L_{\rm crit} \approx 1.5708.
\end{equation}
For $L > L_{\rm crit}$ there is no connected entangling surface but the disconnected entangling surface consisting of two components $x=-L/2$ and $x= L/2$ still exists. For the latter one can straightforwardly calculate the renormalized entanglement entropy as 
\begin{equation}
S_{\rm ren} = - \frac{V_2 \sigma^2}{12 G_5}. 
\end{equation}
The renormalized entanglement entropy is plotted in Figure~\ref{nc:fig:one}: its first derivative is discontinuous at $L = L_{\rm crit}$. For small values of $L$, the analytic expressions~\eqref{nc:analytic} is valid:
\begin{equation}
S_{\rm ren} =  \frac{V_2}{G_5} \left ( - {\left ( \frac{2 \sqrt{\pi} \Gamma( \frac{2}{3}) }{\Gamma (\frac{1}{6})} \right )}^2  \frac{1}{8 L^2} + C_1 \sigma^4 L^2 + \cdots \right ) \label{nc:eq:disk-small-sigma-fit}
\end{equation} 
and the constant $C_1$ can be determined as:
\begin{equation}
  C_1 \approx -0.03137. \label{nc:eq:disk-small-sigma-coeff}
\end{equation}

\FIGURE[t]{\includegraphics*[width=\linewidth]{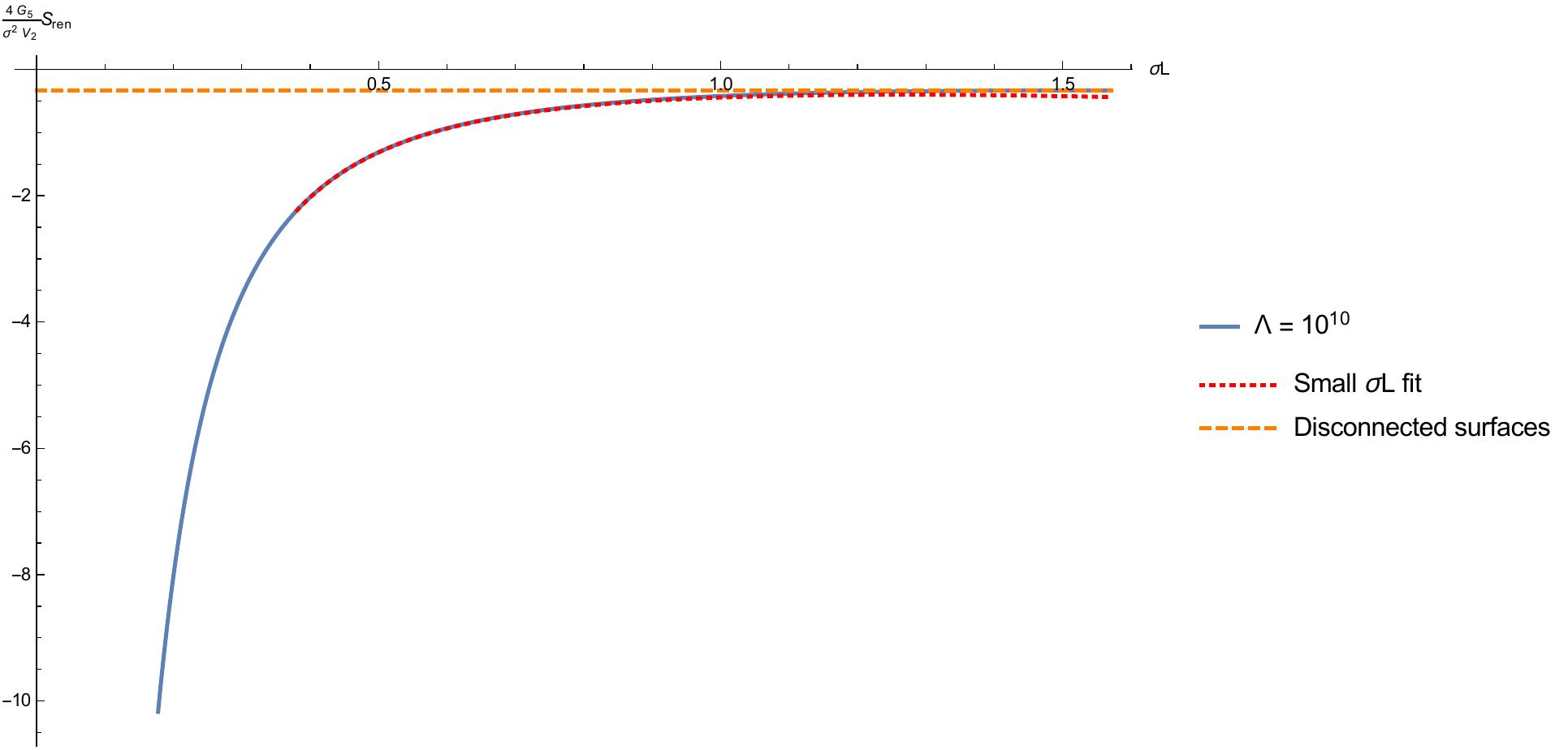}
  \caption{The renormalized entropy for Coulomb branch disk distribution. The blue line shows the numerical results for a cut-off of $\Lambda = 10^{10}$, the dotted red line shows the small $\sigma L$ fit of equation~\eqref{nc:eq:disk-small-sigma-fit}, the dashed yellow line shows the value of the renormalized entropy for disconnected surfaces.}  \label{nc:fig:one}}


\subsubsection{Coulomb branch spherical distribution}

We now consider the renormalized entanglement entropy of slab domains on the Coul\-omb branch of ${\cal N} = 4$ SYM for the case of a spherical distribution of branes, preserving $SO(4) \times SO(2)$ symmetry. The equations of motion follow from~\eqref{nc:bact1}, with the superpotential being \cite{Freedman:1999gk}
\begin{equation}
W(\phi) = \frac{2}{3} \exp \left ( -\frac{2 \phi}{\sqrt{6}} \right ) + \frac{1}{3} \exp \left (  \frac{4 \phi}{\sqrt{6}} \right )
\end{equation}
The metric in five-dimensional gauged supergravity is then
\begin{equation}
ds^2 = \lambda^2 w^2 \left ( \frac{d w^2}{\lambda^6 w^4} + dx \cdot dx \right )
\end{equation}
with 
\begin{equation}
\lambda^6 = \left ( 1 - \frac{\sigma^2}{w^2} \right ).
\end{equation}
Here the coordinate $w \rightarrow \infty$ at the conformal boundary and $\sigma$ characterises the expectation value of the dual scalar operator. The scalar field can be expressed by the relation
\begin{equation}
\sigma^2 \frac{ e^{ -\frac{2}{\sqrt{6}} \phi}}{1 - e^{-\sqrt{6} \phi}} =  - \lambda^2 w^2. 
\end{equation}
Using the standard Fefferman-Graham coordinates near the conformal boundary:
\begin{align}
ds^2 &= \frac{1}{\rho^2} d \rho^2 + \frac{1}{\rho^2} \left ( 1 - \frac{1}{18} \sigma^4 \rho^4 + \cdots   \right ) dx \cdot dx \\
\phi &= \frac{1}{\sqrt{6}} \sigma^2 \rho^2 + \cdots \nonumber
\end{align}
We can then read off the expectation values of the dual stress energy tensor and scalar operator, following~\cite{Bianchi:2001de,Bianchi:2001kw}:
\begin{equation}
\langle T_{ij} \rangle = 0 
\qquad
\langle {\cal O} \rangle = \frac{N^2}{\sqrt{6} \pi^2} \sigma^2
\end{equation}
where we use the standard relation between the Newton constant and the rank of the dual gauge theory:
\begin{equation} 
\frac{1}{16 \pi G_5} = \frac{N^2}{2\pi^2}.
\end{equation}
The vanishing of the dual stress energy tensor is required given the supersymmetry but again careful holographic renormalization is required to derive this answer. 

\bigskip

The regulated entanglement entropy is then 
\begin{equation}
S_{\rm reg} = \frac{V_2}{2 G_5} \int_{w_0}^{\Lambda} dw \frac{ w^3 \sqrt{w^2 - \sigma^2}}{\sqrt{w^4 ( w^2 - \sigma^2 ) - c^2}}
\end{equation}
and the required counterterm is expressed in terms of the regulated area of the boundary of the entangling surface i.e.\ there are counterterm contributions
\begin{equation}
S_{\rm ct} = - \frac{V_2}{8 G_5} \Lambda^2 {\left ( 1 - \frac{\sigma^2}{\Lambda^2} \right )}^{\frac{1}{3}}
\end{equation}
at each side of the slab. (The total contribution is therefore twice this value.) Note that the counterterms in this case clearly contribute both divergent and finite parts: expanding in powers of the cutoff $\Lambda$
\begin{equation}
S_{\rm ct} = - \frac{V_2}{8 G_5} \Lambda^2 + \frac{V_2}{24 G_5} \sigma^2 + \cdots 
\end{equation}
It is then convenient to write the entanglement entropy  in terms of dimensionless quantities as 
\begin{equation}
S_{\rm ren} = \frac{V_2 \sigma^2}{4 G_5}  \lim_{\tilde{\Lambda} \rightarrow \infty} \left ( \int_{y_0}^{\tilde{\Lambda} } dy \frac{ y \sqrt{y^2 -1}}{\sqrt{y^2 ( y - 1 ) - y_0^2 (y_0 -1)}} - \tilde{\Lambda} + \frac{1}{3} \right ),
\end{equation}
where $\tilde{\Lambda}$ is a rescaled dimensionless cutoff. Implicitly this expression assumes that $\sigma^2 \neq 0$ and $y_0$ is the turning point of the surface. Then
\begin{equation}
\sigma L = y_0 \sqrt{y_0 - 1}  \int_{y_0}^{\infty} \frac{dy}{y \sqrt{(y-1)} \sqrt{y^2 (y-1) - y_0^2 (y_0 - 1)}}
\end{equation}
These integrals can again be computed numerically. As in the previous case, for fixed $\sigma$ there is a maximal value of $L$ for which a connected entangling surface exists. The critical value is 
\begin{equation}
\sigma L_{\rm crit} \approx 0.8317
\end{equation}
For lengths grater than the critical length, the minimal surface is disconnected and the renormalized entanglement entropy can be calculated analytically to give
\begin{equation}
S_{\rm ren}  = - \frac{V_2 \sigma^2}{6 G_5}. 
\end{equation}
For subcritical values, there are two possible surfaces with turning points $y_0$ for each width $L$ and one must choose the surface for which the renormalized area is minimised. 

The renormalized entanglement entropy is plotted in Figure~\ref{nc:fig:two}. There is a phase transition between the connected and disconnected entangling surfaces at
$L_c$ such that  
\begin{equation}
\sigma L_c \approx 0.75
\end{equation}
i.e. $L_c < L_{\rm crit}$, and the entanglement entropy is saturated for $L \ge L_c$.  In the regime of small $L$ 
the analytic expressions~\eqref{nc:analytic} are valid:
\begin{equation}
S_{\rm ren} =  \frac{V_2}{G_5} \left ( - {\left( \frac{2 \sqrt{\pi} \Gamma( \frac{2}{3}) }{\Gamma (\frac{1}{6})} \right )}^2  \frac{1}{8 L^2} + C_1 \sigma^4 L^2 + \cdots \right )\label{nc:eq:sphere-small-sigma-fit}
\end{equation} 
where the constant $C_1$ can be determined as:
\begin{equation}
  C_1 \approx -0.03167
\end{equation}

\FIGURE[t]{\includegraphics*[width=\linewidth]{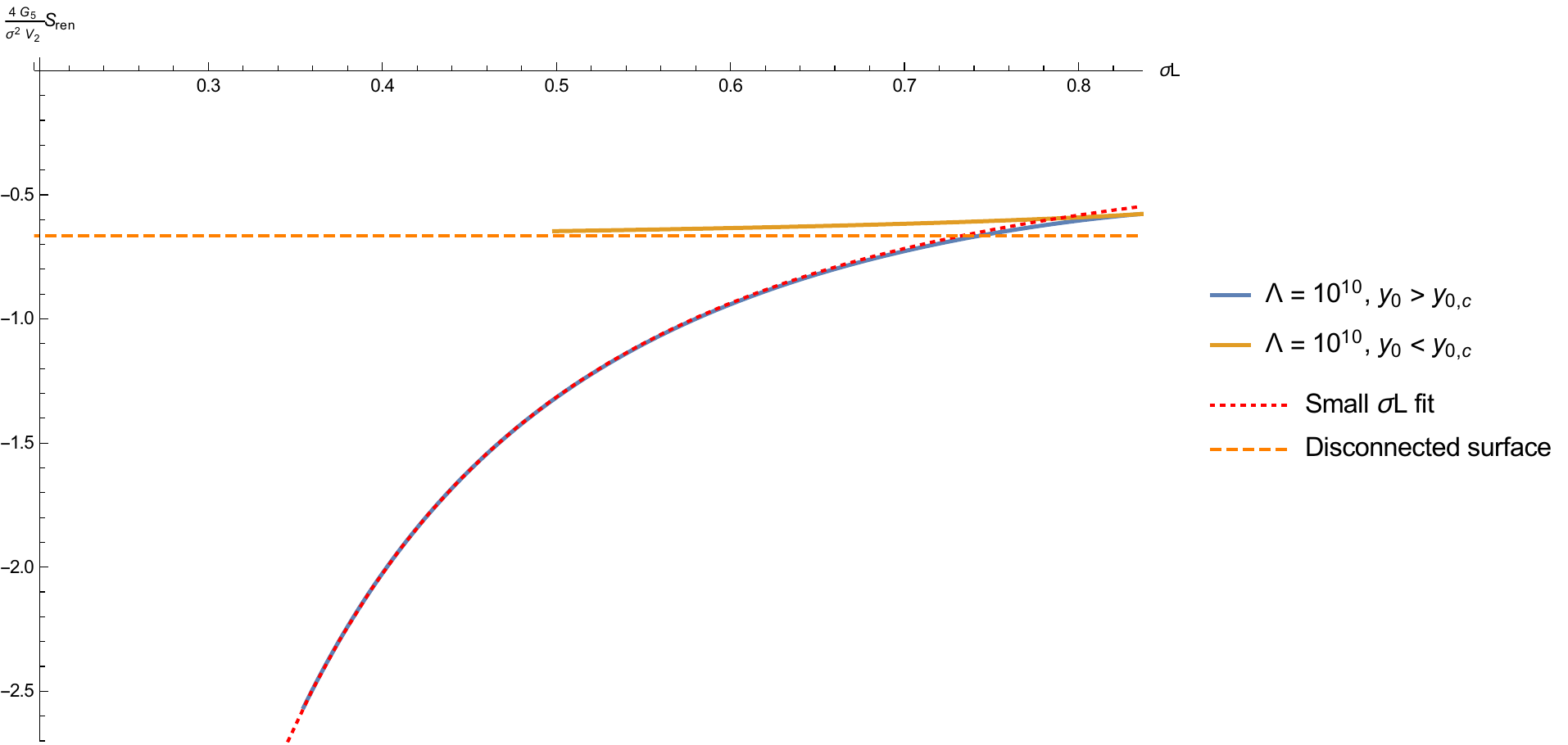}
\caption{The renormalized entropy for Coulomb branch sphere distribution. The solid blue and solid orange lines indicate the renormalized entanglement entropy for the two possible connected minimal surfaces. The dashed orange line indicates the entanglement entropy for the disconnected surface. The dotted red line shows the small $\sigma L$ fit of equation~\eqref{nc:eq:sphere-small-sigma-fit}.}
\label{nc:fig:two}}

\subsection{Operator driven RG flow}

In this section we consider the case of an operator driven RG flow, the GPPZ flow~\cite{Girardello:1999bd}. 
The equations of motion again follow from~\eqref{nc:bact1}, with the superpotential being
\begin{equation}
W(\phi) = \frac{1}{2}  \left ( 1 + \cosh \left  ( \frac{2 \phi}{\sqrt{3}} \right ) \right )
\end{equation}
The metric can be expressed as 
\begin{equation}
ds^2 =  \frac{d \rho^2}{\rho^2} + \frac{1}{\rho^2} (1 - \mu^2 \rho^2) dx \cdot dx 
\end{equation}
while the scalar field is given by 
\begin{equation}
\phi = \frac{\sqrt{3}}{2}  \log \left ( \frac{1 +  \mu \rho}{1 - \mu \rho} \right )
\end{equation}
The scalar $\phi$ is dual to a dimension three operator. By expanding near the conformal boundary and using the holographic renormalization dictionary,~\cite{Bianchi:2001de,Bianchi:2001kw} showed that the GPPZ solution is dual to a deformation (proportional to $\mu$) of ${\cal N} =4$ SYM by the dimension three scalar operator, with the expectation values of the operators being
\begin{equation}
\langle {\cal T}_{ij} \rangle = \langle {\cal O} \rangle = 0. 
\end{equation}
The vanishing stress energy tensor is again required by supersymmetry while the vanishing of the scalar VEV reflects the explicit (as opposed to spontaneous) symmetry breaking.

Now let us consider the renormalized entanglement entropy of a strip region in this geometry. The entanglement entropy can be expressed as 
\begin{equation}
S = \frac{V_2}{2 G_5} \int \dd \rho \frac{(1 - \mu^2 \rho^2)}{\rho^3}\sqrt{1 + (1 - \mu^2 \rho^2){(x')}^2}
\end{equation}
The overall dependence on the deformation $\mu$ can be scaled out to give
\begin{equation}
S = \frac{V_2 \mu^2}{2 G_5} \int \dd v \frac{(1-v^2)}{v^3} \sqrt{1 + (1-v^2){(\partial_v X)}^2}
\end{equation}
where $v = \mu \rho$ and $X = \mu x$. Then the entangling surface of width $L$ satisfies
\begin{equation}
\mu L = 2 \lambda \int_0^{v_0} \dd v \frac{v^3}{\sqrt{{(1 - v^2)}^4 - v^6 \lambda^2 (1 - v^2)}}
\end{equation}
where the integration constant $\lambda$ is related to the turning point of the surface $v_0$ by
\begin{equation}
\lambda = \frac{{(1 - v_0^2)}^{3/2}}{v_0^3}.
\end{equation}
As in the previous cases there is a phase transition between a connected solution for $\mu L < \mu L_{\textrm{crit}}$ and a disconnected solution for $\mu L > \mu L_{\textrm{crit}}$ where
\begin{equation}
  \mu L_{\textrm{crit}} \approx 0.3008.
\end{equation}
The regulated entanglement entropy is then
\begin{equation}
S_{\textrm{reg}} = \frac{V_2 \mu^2}{2 G_5} \int_{\epsilon}^{v_0} \dd v \frac{{(1 - v^2)}^{5/2}}{v^3 \sqrt{{(1-v^2)}^3 - v^6 \lambda^2}}. \label{nc:reg2}
\end{equation}
The counterterms for the entanglement entropy can be derived from the bulk action counterterms using the replica trick:
\begin{equation}
S_{\textrm{ct}} = - \frac{1}{8 G_5} \int_{\partial \Sigma_{\tilde{\epsilon}}} d^2 x \sqrt{\tilde{h}} \left ( 1 + \frac{2}{3} \phi^2 \log ({\epsilon}) \right )
\end{equation}
where the cutoff in the $\rho$ coordinates is $\tilde{\epsilon} = \epsilon/\mu$.  Evaluating this counterterm gives a contribution from each endpoint of the strip:
\begin{equation}
S_{\textrm{ct}} = -\frac{V_2}{8 G_5} \left(\frac{1}{\epsilon^2} - 1 + 2 \log \epsilon\right),
\end{equation}
which indeed matches the regulated divergences of~\eqref{nc:reg2}. Thus the total renormalized entropy is 
\begin{equation}
S_{\textrm{ren}} = \frac{V_2 \mu^2}{4 G_5} \lim_{\epsilon \to 0} \left(2 \int_{\epsilon}^{v_0} \dd v \frac{{(1 - v^2)}^{5/2}}{v^3 \sqrt{{(1-v^2)}^3 - v^6 \lambda^2}} - \frac{1}{2 \epsilon^2} + \frac{1}{2} - \log \epsilon \right).
\end{equation}

These integrals can once again be evaluated numerically, the results of which are plotted in Figure~\ref{nc:fig:gppz-plot}. As in the case of the spherical brane distribution, there are two possible turning points for a given length $L < L_\textrm{crit}$, the branch with $v_0 < v_{0,\textrm{crit}}$ is favoured for all such $L$. Both branches are positive near $L = L_{\textrm{crit}}$, whereas it can be shown analytically that the renormalized entropy is zero for disconnected surfaces and so there is a transition from the connected to disconnected surface solutions at around
\begin{equation}
  \mu L_c \approx 0.27
\end{equation}
where the entanglement entropy has a discontinuous derivative.

\FIGURE[t]{\includegraphics[width=\linewidth]{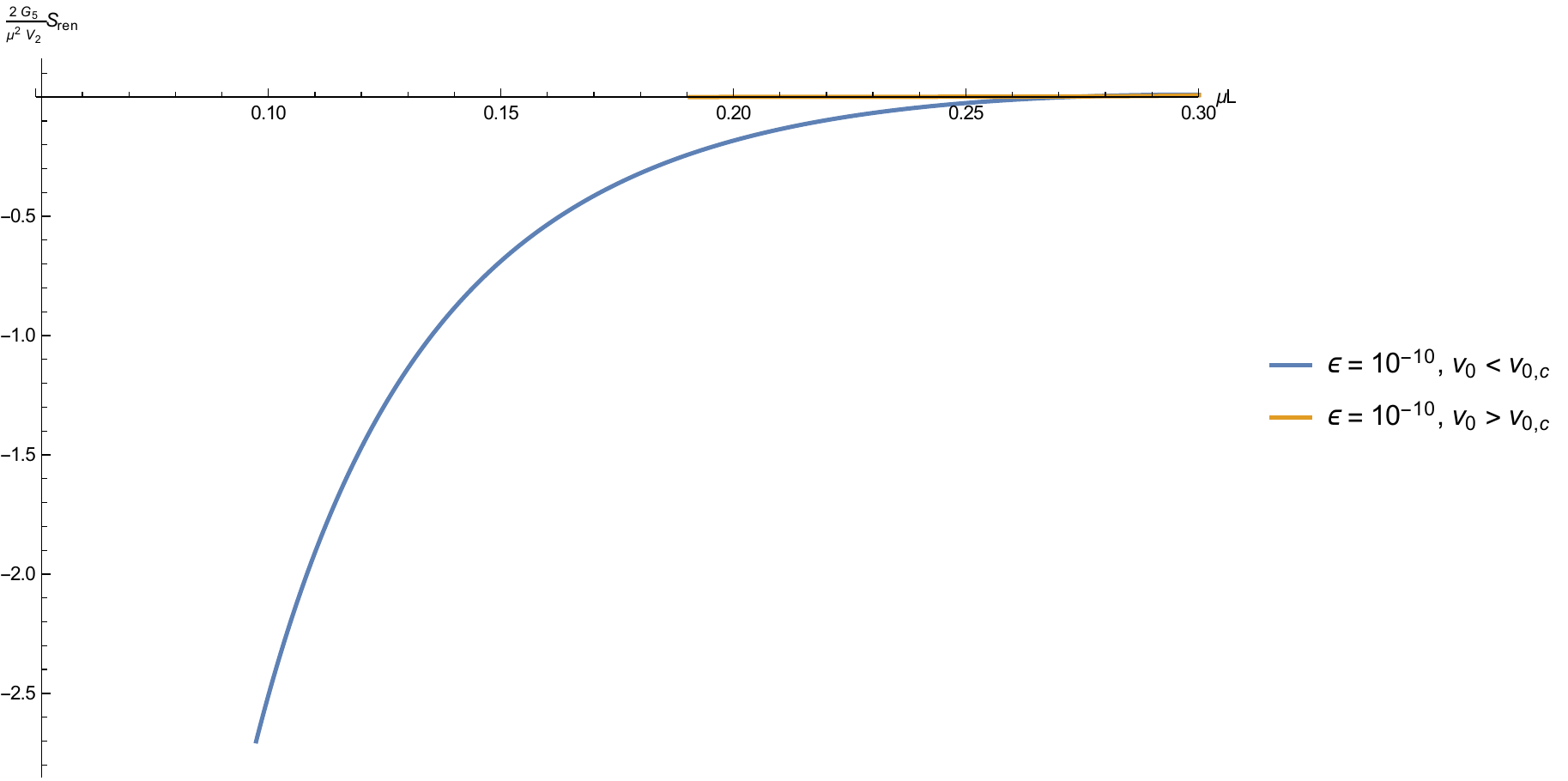}
  \caption{The renormalized entropy for the GPPZ flow. The solid blue line and solid orange lines indicate the renormalized entropy for the two possible connected minimal surfaces. Note that $S_{\textrm{ren}} > 0$ near $\mu L_{\textrm{crit}}$ for the connected solutions.}\label{nc:fig:gppz-plot}}


\section{Non-relativistic deformations} \label{six}

Schr\"{o}dinger metrics in $(p+3)$ dimensions can be written as \cite{Son:2008ye}
\be
ds^2 = \frac{b^2}{r^{2z}} (dx^+)^2 + \frac{1}{r^2} \left ( dr^2 + dx^+ dx^- + dx^i dx_i \right )
\ee
where the index $i$ runs over $p$ directions. The light cone coordinates can be rewritten as
\be
x^{\pm} = (y \pm t)
\ee
The metric can be supported by (real) massive gauge fields provided that $b^2 > 0$ for $ z < 1$ and $b^2 < 0 $ for $z > 1$. 

In both cases the dual field theory can be understood as a deformation of the CFT by an operator that breaks the relativistic symmetry but respects non-relativistic scaling invariance i.e. the dual theory has the form \cite{Guica:2010sw}
\be 
I_{CFT} + \int d^{p+2} x  \; | b | {\cal O}_{-} + \cdots 
\ee
where the operator ${\cal O}_{-}$ is a vector (or tensor) that picks out the $x^+$ direction. The deformation is relevant (dimension less than $(p+2)$) with respect to the original conformal symmetry for $z < 1$ and irrelevant for $z > 1$. 

Let us consider the case of $z < 1$, so that the theory remains UV conformal; this case was explored in detail in \cite{Costa:2010cn}.  
We can then specify a spacelike entangling region in the dual quantum field theory, at constant $t$, defined by $y(x^i)$. (For $z > 1$, the situation is more complicated as surfaces of constant $t$ are not spacelike at infinity and we will not discuss this case further here.) 

We can illustrate the behaviour of the entangling surfaces by two cases: a slab of width $L$ in the $y$ direction and a slab of width $L$ along one of the $x^i$ directions. 
In the latter case the entangling surface in the bulk is described by $w(r)$ (where $w$ is the direction transverse to the entangling region) and the entangling functional is
\be
S_1 =  \frac{R_y V_{p-1}}{2 G} \int  \frac{1}{r^{(p+1)}} \sqrt{1 + b^2  r^{2(1-z)}} \sqrt{1 + w'(r)^2} dr
\ee
where $R_y$ is the regulated length of the $y$ direction and $V_{p-1}$ is the regulated volume of the $x^i$ directions, excluding $w$. 

In the other case the entangling surface is described by $y(r)$ and the entangling functional is 
\be
S_2 = \frac{V_p}{2 G} \int \frac{1}{r^{(p+1)}} \sqrt{1 + (1 + b^2 r^{2(1-z)}) y'(r)^2} dr 
\ee
where now $V_p$ is the regulated volume of the $x^i$ directions. 

Note that both entangling functionals can be expressed in the form 
\be
S = {\cal N} \int f(r) \sqrt{ 1 + g(r) w'(r)^2} dr
\ee
for suitable choices of the overall normalisation ${\cal N}$ and the functions $(f(r),g(r))$. Then the width of the entangling region is given by
\be
L = 2 \int^{\infty}_{r_o} \frac{dr}{\sqrt{ g(r) (g(r)f(r)^2 - g(r_o) f(r_o)^2)}}
\ee
where $r_o$ is the turning point of the minimal surface and 
\be
S = {\cal N} \int_{r_o}^{\infty} \frac{ f(r)^2 \sqrt{g(r)}} {\sqrt{g(r) f(r)^2  - g(r_o) f(r_o)^2}} dr.
\ee
When $f(r)$ and $g(r)$ are monomials of $r$, the renormalized entanglement entropy can be calculated exactly using the AdS$_D$ result derived in section \ref{three}.

\bigskip

For the slab along the $x^i$ directions, the renormalized entanglement entropy interpolates between the AdS$_{p+3}$ result (for small slab widths) 
\be
(S_1)_ {\rm ren} =  - \frac{R_y V_{p-1} }{4 p G L^p } {\left( \frac{ 2 \sqrt{\pi} \Gamma  \left ( \frac{1}{2 (p+1)} + \frac{1}{2} \right )}{\Gamma  \left ( \frac{1}{2 (p+1)} \right )} \right )}^{p+1} \label{UV}
\ee
and the following result for large slab widths
\be
(S_{1})_ {\rm ren} = -  \frac{b^2 R_y V_{p-1}}{4 (p+z-1) G L^{p+z-1}} {\left( \frac{ 2 \sqrt{\pi} \Gamma  \left ( \frac{1}{2 (p+z)} + \frac{1}{2} \right )}{\Gamma  \left ( \frac{1}{2 (p+z)} \right )} \right )}^{p+z}
\ee
The latter expression applies for $b L^{1-z} \gg 1$, in which case the functional is approximated by
\be
S_1 =  \frac{R_y V_{p-1}}{2 G} \int  \frac{b}{r^{(p+z)}} \sqrt{1 + w'(r)^2} dr
\ee
which is precisely the functional analysed in section \ref{three} (taking $D = p + z$). 

\bigskip

In the other case, the renormalized entanglement entropy is also given by the AdS$_{p+3}$ result for small slab widths while at large slab widths $b L^{1-z} \gg 1$ the relevant functional is \be
S_2 = \frac{V_p}{2 G} \int \frac{1}{r^{(p+1)}} \sqrt{1 +  b^2 r^{2(1-z)} y'(r)^2} dr. \label{red-func}
\ee
Consider first the special case of $z=0$. By a change of variable we can express this functional as 
\be
S_2 = \frac{ b V_p}{2 G} \int \frac{1}{w^{\frac{p}{b}}} \sqrt{1 +   \dot{y}(w)^2} dw, 
\ee
where $\dot{y} = dy/dw$. From the general result \eqref{nc:hol-co}, we can now read off the renormalized entanglement entropy as 
\be
(S_2)_{\rm ren} =  - \frac{ b V_{p} }{4 (\frac{p}{b}-1) G L^{\frac{p}{b}-1} } {\left( \frac{ 2 \sqrt{\pi} \Gamma  \left ( \frac{b}{2 p} + \frac{1}{2} \right )}{\Gamma  \left ( \frac{b}{2 p} \right )} \right )}^{\frac{p}{b}}
\ee 
For $0 < z < 1$, the functional \eqref{red-func} can be expressed as 
\be
S_2 =  \frac{ b V_p}{2 G} \left ( \frac{1}{z b} \right )^{\frac{p}{z} + 1} \int \frac{dw}{w^{\frac{p}{z} + 1}} \sqrt{1 +   \dot{y}(w)^2} dw
\ee
and thus from \eqref{nc:hol-co}
\be
(S_2)_{\rm ren} =  - \frac{ b V_{p} }{4 (\frac{p}{z}) G L^{\frac{p}{z}} } \left ( \frac{1}{z b} \right )^{\frac{p}{z} + 1} 
{\left( \frac{ 2 \sqrt{\pi} \Gamma  \left ( \frac{1}{2 (\frac{p}{z} + 1) } + \frac{1}{2} \right )}{\Gamma  \left ( \frac{1}{2 ( \frac{p}{z} +1) } \right )} \right )}^{\frac{p}{z} + 1}
\ee
which is manifestly consistent with \eqref{UV} for $b=z=1$. 

\bigskip

Thus the renormalized entanglement entropy scales differently for large slab widths (such that $b L^{1-z} \gg 1$), depending on the orientation of the slab with respect to the $y$ direction along which the theory is deformed away from conformality. In this case the explicit symmetry breaking is associated with a breaking of the relativistic symmetry, while preserving non-relativistic scale invariance, and the renormalized entanglement entropy does not have a discontinuity in its derivative and does not saturate in the deep IR. (Note however that the 
Schr\"{o}dinger geometry has a null curvature singularity and thus quantum corrections to the geometry may change the deep IR behaviour.)

\section{Interpretation and comparison to QFT results} \label{seven}

In the previous sections we have explored the renormalized entanglement entropy for slab domains in a variety of holographic models. While the general method of area renormalization is applicable to entangling domains of any shape, it is particularly convenient to use slab domains for several reasons. Firstly, the equations of motion admit first integrals, thus simplifying the analysis. Secondly, slab entangling domains have been analysed for a variety of quantum field theories in the literature. Note that the previous literature does not compute the renormalized entanglement entropy, but typically extracts instead
\be
c(L) = \frac{\partial S}{\partial L}.
\ee 
As we discussed in the introduction, in any local quantum field theory the divergences in the entanglement entropy are necessarily independent of the width of the slab, $L$, and thus $c(L)$ is manifestly UV finite. 

Consider now the renormalized entanglement entropy. The counterterms include finite contributions, as illustrated in the previous section, but these finite contributions are independent of the width of the slab, as the counterterms are expressed in terms of local quantities at the boundaries of the entangling region. Therefore 
\be
\frac{\partial S_{\rm ren}}{\partial L} = \frac{\partial S_{\rm reg}}{\partial L} = c(L),
\ee
and thus the slope of our renormalized quantity matches the $c$ function defined in earlier literature. This statement can be expressed as
\be
S_{\rm ren} = \int c (L) dL +  s_0 \label{generic1}
\ee
where $s_0$ is independent of $L$, but dependent on parameters of the theory. Thus the renormalized quantity is an integrated version of the $c$ function. (Note that the $c(L)$ is defined in various different ways in the literature. For example, \cite{Nishioka:2006gr} use a definition of the entropic $c$ function that incorporates factors of $L$.)

\bigskip

Next let us consider the UV and IR behaviour of the renormalized entanglement entropy. The renormalized entanglement entropy measures the residual entanglement between the entangling region and its complement, after subtracting the divergent contributions arising from entanglement at the boundary. In the ground state of a conformal field theory, correlation functions are characterised by power law behaviour and thus it would be reasonable to expect that the residual entanglement scales inversely with the width of the slab entangling region (and extensively with the length of the slab region). 

This heuristic argument is in agreement with the explicit holographic result \eqref{nc:hol-co}. For a slab region in the ground state of a conformal field theory, the $L$ independent contribution $s_0$ in \eqref{generic1} is necessarily zero, as there is no dimensionless ratio that is independent of $L$. The renormalized entanglement entropy is thus determined entirely by the $c$ function, with the positivity of the latter implying the negativity of the former.  For non-conformal branes, the entanglement entropy is controlled by the conformal structure in $(d - 2 \alpha \gamma)$ dimensions, and therefore similar arguments apply.

\bigskip

Suppose that in the IR of the theory correlation functions fall off exponentially with characteristic mass scale $\sigma$; entanglement is thus significant only on length scales of order $\sigma^{-1}$ from the entangling region boundaries. If the width of the entangling region $L$ is much greater than this length scale, then we would expect the renormalized entanglement entropy to saturate to a value that is independent of $L$. On dimensional grounds this residual entanglement entropy must then scale for a $d$-dimensional theory as $V_{d-2} \sigma^{d-2}$ for a slab region of area $V_2$. Thus there is a non-vanishing constant term in \eqref{generic1} which would not be seen in the $c$ function (which is in such cases zero for large $L$). 

This behaviour can be seen in a number of explicit QFT calculations. In \cite{Hertzberg:2010uv} the entanglement entropy for massive scalar fields in various dimensions was computed, and expressed in term of the derivative of the entanglement entropy with respect to the mass $\mu$. The latter is sensitive to the contributions to the renormalized entanglement entropy that are independent of $L$, and hence are lost from the $c$ function. For example, for $d=3$, it was shown that 
\be
S_{\mu} \equiv - \mu^2 \frac{\partial S}{\partial \mu^2} = \frac{ V_1 \mu}{24}
\ee
with $V_1 \gg \mu^{-1}$ the regulated length of the slab region. Integrating this expression results in a finite contribution to the renormalized entanglement entropy 
\be
S_{\rm ren} = - \frac{V_1 \mu}{12},
\ee
in agreement with the above arguments.  

In $d=4$ the analogous expression is
\be
S_{\mu} \equiv  \mu^4 \frac{\partial^2 S}{\partial (\mu^2)^2}= \frac{V_2\mu^2}{48 \pi}.
\ee
These terms arise from logarithmic divergences in the regulated entanglement entropy
\be
S = \cdots + \frac{V_2}{48 \pi} \left ( \mu^2 \log (\mu^2 \epsilon^2) - \mu^2 \right ). 
\ee
Whenever there is a logarithmic divergence, the 
renormalized entanglement entropy has scheme dependence, corresponding to the choice of finite counterterms \cite{Taylor:2016aoi}. 

Such logarithmic divergences occur in particular for CFTs in $d$ dimensions deformed by operators of dimension $\Delta = d/2 + 1$ \cite{Rosenhaus:2014zza,Jones2015}. The logarithmic divergences are removed by counterterms
of the form
\be
\int_{\partial \Sigma} d^{d-2} x \sqrt{\gamma} \phi^2 \log \epsilon
\ee
in holographic realisations, where $\phi$ is the scalar field dual to the deforming operator. By rescaling $\epsilon \rightarrow e^\alpha \epsilon$ this counterterm will change to
\be
\int_{\partial \Sigma} d^{d-2} x \sqrt{\gamma} \phi^2 ( \log \epsilon + \alpha)
\ee
with the latter term being finite (due to the operator dimension).  In particular, using the leading asymptotic behaviour for the scalar field, the latter term contributes a term
\be
\alpha V_{d-2} \phi_s^2
\ee
to the renormalized entanglement entropy, where $\phi_s$ is the operator source. Therefore, the renormalized entanglement entropy depends explicitly on 
the choice of the coefficient of the finite term $\alpha$. (More generally, operators of dimension $\Delta = d (1 - 1/2n) + 1/n$ are associated with logarithmic divergences \cite{Taylor:2016aoi} and hence lead to finite terms in the entanglement entropy behaving as $V_{d-2} \phi_s^{2n}$.) 

In supersymmetric theories, the ambiguity can be fixed by requiring that the renormalization scheme for the partition function respects supersymmetry and then using the replica trick to derive the counterterms for the entanglement entropy. The renormalization scheme for GPPZ, which indeed corresponds to a CFT deformed by a supersymmetric operator of dimension $\Delta = d/2 + 1$, was constructed to respect supersymmetry \cite{Bianchi:2001de,Bianchi:2001kw}. It is thus perhaps unsurprising that the supersymmetric scheme implies that the renormalized entanglement entropy in this case vanishes in the deep IR. 

\bigskip

The discontinuity in the derivative of the entanglement entropy with respect to $L$ in a holographic confining theory was first described in  \cite{Klebanov:2007ws}. In the examples of explicit and spontaneous symmetry discussed here the renormalized entanglement entropy always saturates in the IR, and there is a discontinuity in the derivative of the entanglement entropy at the critical value of $L$, at which the dominant entangling surface becomes disconnected. Note however that the slope of the derivative can be small close to the transition point, as in one of our Coulomb branch examples, and one thus needs to ensure that the numerical resolution is sufficient to capture the discontinuity in the derivative. 

\bigskip

In addition to calculations of the entanglement entropy in free field theories, various calculations of the entanglement entropy for slab regions have been carried out in lattice gauge theories. In \cite{Buividovich:2008kq} the entanglement entropy for a slab of width $L$ in a four-dimensional $SU(2)$ gauge theory was studied numerically. The results of this study are in agreement with the behaviour found here. The derivative of the entanglement entropy with respect to $L$ has a discontinuity at a critical value, as found in holographic confining theories in \cite{Klebanov:2007ws} and discussed above, and it was also observed that there are finite contributions to the entanglement entropy which scale as $1/L^2$ for small width entangling regions. 

While \cite{Buividovich:2008kq} did not extract the renormalized entanglement entropy, their results imply that the renormalized entanglement entropy would scale as $1/L^2$ for small width entangling regions. The $SU(2)$ gauge theory is asymptotically free and thus one would expect the renormalized entanglement entropy for small regions to be captured by free gluons, which indeed scales in accordance with the conformal result discussed earlier in the paper. Note that the residual finite contributions at large $L$ were not computed in  \cite{Buividovich:2008kq}.

A more recent lattice simulation \cite{Itou:2015cyu} studied entanglement entropy for slab regions in $SU(3)$ gauge theory in four dimensions. The generic features are similar to those found in the SU(2) theory (free at small distances, $c(L)$ goes to zero at finite $L$), although the detailed features near the critical length differ between $SU(2)$ and $SU(3)$. In particular, $c(L)$ seems to go smoothly to zero at the critical length, and therefore there is no discontinuity in the derivative of the entanglement entropy with respect to $L$. 
As in  \cite{Buividovich:2008kq}, only the vanishing of the derivative of the renormalized entanglement entropy for large $L$ was shown; the residual finite entanglement entropy was not computed.

\section{Conclusions and outlook} \label{eight}

In this paper we have explored renormalized entanglement entropy for slab domains, for a variety of different holographic theories. We have shown that the renormalized entanglement entropy captures not just the features of the previously discussed entangling $c$ function, but also the deep IR behaviour of symmetry breaking theories (where the $c$ function vanishes). It would be interesting to analyse the properties of renormalized entanglement entropy for other common entangling regions, such as spheres and hypercubes. Note however that the latter are considerably more complicated to compute holographically: the equations of motion for the minimal surfaces do not admit first integrals and the vertices of hypercubes are generally associated with additional logarithmic counterterms in the entanglement entropy. 

The examples discussed in this paper indicate the existence of general bounds on the renormalized entanglement entropy: $S_{\rm ren} \le 0$ with $S_{\rm ren} \rightarrow 0$ for supersymmetric RG flows triggered by operator deformations. It would be interesting to develop proofs of these bounds in future work. Related bounds were discussed in \cite{Fischetti:2016fbh}, although the functional analysed in \cite{Fischetti:2016fbh} is not identical to the renormalized entanglement entropy considered here. 

Note that there are heuristic arguments why $S_{\rm ren} \le 0$. For CFTs in odd dimensions, following \cite{Casini2011}, the renormalized entanglement entropy for spherical regions is related to the partition function on a sphere, and the negativity of the renormalized entanglement entropy is thus related to the conjectured positivity of the F quantity \cite{Jafferis:2011zi}. 

More generally, the renormalized entanglement entropy coincides with minus the (renormalized) Euclidean action for a D$(d-1)$-brane with no worldvolume gauge fields and no Chern-Simons couplings to background fluxes i.e. the latter is also a minimal surface. The (renormalized) Euclidean action is positive semi-definite for stable D-brane embeddings, and vanishes for supersymmetric D-brane embeddings. This heuristic argument suggests that the renormalized entanglement entropy should be negative semi-definite but 
does not however explain why the renormalized entanglement entropy is zero  in the IR for supersymmetric operator driven flows but not for supersymmetric VEV driven flows. 

Holography allows us to explore entanglement entropy for a wide variety of strongly coupled quantum field theories. In this work we have extracted from existing perturbative and lattice results the behaviour of the renormalized entanglement entropy for slabs but it would clearly be interesting to explore renormalized entanglement entropy directly within perturbative quantum field theory, using varied renormalization methods. The replica trick allows us to derive the counterterms for the entanglement entropy but it would be useful to understand the role of these counterterms in computations of renormalized entanglement entropy via twist field correlators. 

There has been considerable progress in understanding the computation of entanglement entropy in lattice gauge theories, see for example \cite{Buividovich:2008kq,Buividovich:2008yv,Aoki:2015bsa,Itou:2015cyu,Aoki:2016lma}, and it would be interesting to explore how the continuum limit of such computations can be matched with our definition of renormalized entanglement entropy. 

More generally, one would hope that it may become possible to calculate entanglement entropy for certain supersymmetric theories on the lattice in the near future - see for example \cite{Catterall:2014mha} for recent progress on simulating ${\cal N} = 4$ SYM. We can rewrite the holographic result \eqref{nc:hol-co} for the renormalized entanglement entropy for a slab in ${\cal N} = 4$ SYM as
\be
S_{\rm ren} \approx - 0.114 \frac{N^2 V_y}{L^2}.
\ee
Conformal invariance implies that the renormalized entanglement entropy has a leading behaviour at large $N$
\be
S_{\rm ren} = - f(g_{\rm YM}^2 N) \frac{N^2 V_y}{L^2}
\ee
where $f(g_{\rm YM}^2 N)$ is a positive function of the 't Hooft coupling; it is this function that one would like to compute perturbatively using lattice simulations. One can estimate the free field value of this function by summing contributions from the six real scalars, four Weyl fermions (equivalent to two Dirac fermions) and the gauge field of ${\cal N} = 4$ SYM. Estimating the gauge field contributions by scaling the recent $SU(3)$ result of \cite{Itou:2015cyu} and taking the other contributions from \cite{Nishioka:2006gr,Casini:2009sr} we obtain $f \sim -0.05$ at zero coupling. This suggests that the magnitude of $f$ increases with the 't Hooft coupling, as one might expect.

\section*{Acknowledgements}

We would like to thank Antonio Rago for useful comments regarding lattice calculations of entanglement entropy. 
This work was supported by the Science and Technology Facilities Council (Consolidated Grant ``Exploring the Limits of the Standard Model and Beyond'').
We thank the Simons Center and the GGI for partial support during the completion of this work. This project has received funding from the European Union's Horizon 2020
research and innovation programme under the Marie Sklodowska-Curie grant
agreement No 690575.

\bibliographystyle{jhep2}

\providecommand{\href}[2]{#2}\begingroup\raggedright\endgroup

\end{document}